\documentstyle[11pt,a4,cite,epsf]{article}
\newcommand{\be}{\begin{equation}}
\newcommand{\ee}{\end{equation}}
\newcommand{\bea}{\begin{eqnarray}}
\newcommand{\eea}{\end{eqnarray}}
\newcommand{\noi}{\noindent}

\newcommand{\cO}{{\cal O}}

\newcommand{\eff}{\mbox{\rm eff}}

\newcommand{\ra}{\rightarrow}
 
\begin{document}
 
\begin{titlepage}
\begin{flushright} CPT-96/P.3324\\ UAB-FT-368\\ 
\end{flushright}
\vspace*{2cm}
\begin{center} {\Large \bf On Renormalons and Landau Poles in Gauge
Field Theories}\\[1.5cm] {\large {\bf Santiago
Peris}$^{a}$\footnote{Work partially supported by
research project CICYT-AEN95-0882.}, and 
{\bf Eduardo de Rafael}$^b$}\\[1cm]

$^a$ Grup de F\'{\i}sica Te\'orica and IFAE\\ Universitat Aut\'onoma
de Barcelona, 08193 Barcelona, Spain.\\[0.5cm] and\\[0.5cm]
$^b$  Centre  de Physique Th\'eorique\\
       CNRS-Luminy, Case 907\\
    F-13288 Marseille Cedex 9, France\\
\end{center}

\vspace*{1.5cm}
\begin{abstract}

It is shown that the commonly accepted relationship between the Landau
singularity in the running coupling constant of QED or QCD and the
renormalon singularities in the Borel sums of perturbation theory
expansions is only a particular feature of the restriction of the
perturbative $\beta$--function to the one loop level.   

\end{abstract}

\vfill

\begin{flushleft} hep-ph/9603359\\
\end{flushleft}
 
\begin{flushleft} March 1996\\
\end{flushleft} \end{titlepage}


{\bf 1.} The success of the Standard Model in describing particle
physics at present energies relies heavily on the perturbation theory
approach as applied to the underlying gauge field theories~:
$SU(3)\times SU(2)_{L}\times U(1)$. Yet, it is known since the early
work of Dyson~\cite{Dy52}  that perturbation theory itself suffers from
ambiguities which originate in the fact that physical quantities are
not analytic in the coupling constants which define the expansion
parameters. This is reflected in a characteristic $k!$--growth pattern
in the coefficients having the same sign of the large order terms in the 
perturbation
series~\footnote{For a comprehensive review of the subject and a
collection of early articles see ref. \cite{LeGZJ90}}. In theories like
QED this growth originates in the large momentum integration region of
virtual photons dressed with vacuum polarization corrections and leads
to singularities in the associated Borel plane; the so--called 
UV--renormalons~\cite{tH78,La77,Pa77,Co81,BS95}. In QCD, it is the low
momentum integration region of virtual gluons dressed with running
couplings which leads to non integrable singularities in the Borel
plane the so--called IR--renormalons~\cite{tH78,Pa79,Mu85}.

The study of renormalon properties in gauge theories is at
present an active field of research. The r\^ole of IR--renormalons
in the operator product expansion of two point functions and their
relationship with non--perturbative inverse power corrections has been
extensively discussed in the
literature~\cite{Da82,Da84,NSVZ84,NSVZ85,Mu85,Da86,Mu93,Sj95}. This and
further discussions which originated after ref.~\cite{BY92} appeared,
has led to a new point of view~\cite{Za92} concerning renormalons
in QCD which focuses on the possibility that their systematic study in a
given hadronic process might suggest generic non--perturbative effects
of a universal nature. This applies to the case of IR--renormalons as
well as to the much less explored r\^ole of
UV--renormalons~\cite{VZ94a,VZ94b} in QCD.

Practically all work on renormalons in the literature is restricted to
the effect of the first coefficient $\beta_{1}$ of the
$\beta$--function. There is a good reason for that~: the position of the
singularities for positive Borel variable is governed by $\beta_{1}$; and  it
is the sign of
$\beta_{1}$ which fully dictates their fate as UV--like
$(\beta_{1}>0)\,$ versus IR--like singularities $(\beta_{1}<0)\,$. It is
also often stated that {\it these singularities are due to the Landau
pole in the running coupling constant}. The purpose of this note is to
discuss some interesting
properties which appear when the first two terms of the $\beta$--function
are fully taken
into account, and to show how they clash with this common belief that 
{\it renormalon singularities are due to the Landau pole in the running
coupling constant.}

\vspace{7 mm}  
{\bf 2.} We shall first review the usual analysis of renormalon
effects in the precise case of the anomalous magnetic moment of
the electron
$a_e$ in QED, the same  example that Lautrup considered in
ref.~\cite{La77}.
The relevant Feynman diagrams are those generated by the
one--renormalon chain in Fig.~1, and their contribution to $a_e$ is
given by the integral

\be \label{eq:ane}
a_{e}=\frac{1}{\pi}\int_{0}^{1}dx (1-x)\,
\alpha_{\eff}\left(\frac{x^2}{1-x},\alpha\right)\,,
\ee
\noi
where $x$ is the Feynman parameter which combines the propagators in
the loop vertex of Fig.~1, and $\alpha_{\eff}$ the QED effective
charge. For large values of the euclidean momentum $k^2$
($k^2/m^2=\frac{x^2}{1-x}$ in our case, with $m$ the electron
mass) the effective charge obeys the renormalization group equation~:

\be \label{eq:rge}
\left( m\frac{\partial}{\partial m} +
\beta(\alpha)\alpha\frac{\partial}{\partial \alpha}\right)
\alpha_{\eff}^{(\infty)}\left(\frac{k^2}{m^2},\alpha\right)=0\, ,
\ee
where the infinity superscript refers to this asymptotic limit for $k^2$.

Equation (\ref{eq:ane}) is an exact integral representation of the
contribution to the electron anomaly from the infinite class of
diagrams which Fig.~1 represents. In particular, if one takes
$\alpha_{\eff}=\alpha\simeq 1/137$ one of course re-obtains the value
$a_{e}=\alpha/2\pi$. When  the number of vacuum
polarization bubbles is large, the integral is then dominated
by the $x\ra 1$ region; i.e. large values of the virtual euclidean loop
momentum, and becomes equivalent to the integral

\be\label{eq:momentum}
a_{e}\simeq
\frac{1}{\pi}\int_{m^2}^{\infty}\frac{dk^2}{k^2}
\left(\frac{m^2}{k^2}\right)^2
\alpha_{\eff}^{(\infty)}\left(\frac{k^2}{m^2},\alpha\right)\,.
\ee
\noi
The insertion in the integrand above of the one loop solution to the
renormalization group equation in (\ref{eq:rge}) with
$\beta(\alpha)=[\beta_{1}=\frac{2}{3}]\frac{\alpha}{\pi}$; i.e., the
insertion of the expression

\be\label{eq:ech1}
\frac{\alpha}{\alpha_{\eff}^{(\infty)}\left(\frac{k^2}{m^2},
\alpha\right)}=1-\frac{\beta_{1}}{2\pi}\alpha\log\frac{k^2}{m^2}
\ee
\noi
in (\ref{eq:momentum}), and the change of variables~:

\be
B\frac{z}{2}=1-\frac{\alpha}{\alpha_{\eff}^{(\infty)}
\left(\frac{k^2}{m^2},\alpha\right)}\,, \qquad \mbox{\rm where} \qquad
B\equiv\frac{\beta_{1}}{2\pi}\,,
\ee
\noi
leads then to the Borel sum~\cite{La77}

\be \label{eq:borel1}
a_{e}=\frac{1}{\pi}\int_{0}^{\infty}dz\,e^{-\frac{z}{\alpha}}
\frac{1}{2-Bz}\,.
\ee
\noi

The singularity (UV renormalon) at $z=2/B$ reminds us that we only know the
function for $z<2/B$, just as we only know
$\alpha_{\eff}^{(\infty)}\left(\frac{k^2}{m^2},\alpha(m)\right)$ in Eq.
(\ref{eq:ech1}) up to a momentum
$k^2=\Lambda_L^2=\exp\left(\frac{1}{B\alpha}\right)$. At this momentum (the Landau scale) the effective charge becomes
infinity. The association of the UV-renormalon ambiguity with the Landau
pole comes from this observation.

Notice that in Eq.~(\ref{eq:borel1}) 
the leading UV renormalon at $z=1/B$ does not contribute, so that the
ambiguity starts at $z\geq 2/B$. The
reason for it is the factor $1-x$ in the numerator in
eq.~(\ref{eq:ane}), which appears because of the electron helicity
conservation of the electromagnetic interaction, and
leads to 2--powers of the factor $m^2/k^2$ in the euclidean momentum
version of the same integral in (\ref{eq:momentum}). In the language of
UV--effective operators of Parisi~\cite{Pa79}, this corresponds to the
tree level insertion of the operator $\left(m^3/\Lambda^4_L\right)
\bar{\psi}(x)\sigma_{\mu\nu} \psi(x) F^{\mu\nu}(x)$.

    
\vspace*{7mm}

{\bf 3.} Let us now study the same $a_e$ observable but in the
presence of the first two terms of the QED
$\beta$--function~:

\be
\beta(\alpha)=[\beta_{1}=\frac{2}{3}]\frac{\alpha}{\pi}+
[\beta_{2}=\frac{1}{2}]\left(\frac{\alpha}{\pi}\right)^2 \,.
\ee
In this equation we have no loss of generality since any $\beta-$function 
can be brought into this form by a (perturbative) coupling constant 
redefinition.

To simplify the notation we shall denote   
\be
\alpha(k)\equiv \alpha_{\eff}^{(\infty)}\left(\frac{k^2}{m^2},
\alpha\right)Ê\qquad \mbox{\rm and}\qquad \delta
\equiv 2\frac{\beta_2}{\beta_{1}^{2}}\,.
\ee
\noi
The general solution of eq.~(\ref{eq:rge}) is then

\be\label{eq:rgegs}
\frac{\alpha(m)}{\alpha(k)}=1-B\alpha(m)\log\frac{k^2}{m^2}+\delta
B\alpha(m)
\left[\log\frac{\alpha(m)}{\alpha(k)}+\log\frac{1+\delta B\alpha(k)}
{1+\delta B\alpha(m)}\right]\,,
\ee
\noi
where $\alpha(m)$ is the boundary condition at $k^2=m^2$. The
shape of these solutions in the plane $\log\frac{k^2}{m^2}$,
$\alpha(k)$ is shown in Fig.~2. In full generality, and depending on
whether the boundary condition is chosen to be $\alpha(m)\geq 0$;
$-\frac{1}{\delta B}\leq\alpha(m)\leq 0$; or $\alpha(m)\leq
-\frac{1}{\delta B}$ the solutions lie along the curves shown in the
regions I, II, or III of Fig.~2. These regions correspond to the
analogous ones in the $\beta$--function of Fig.~3 where the arrows
show the flow of the effective charge as $k^2\ra\infty$. The three
regions are separated by the fixed points at $\alpha^{*}=0$ and
$\alpha^{**}=-\frac{1}{\delta B}$.

The physical region is clearly the one in I. For a given initial
condition $\alpha(m)>0$, the running coupling $\alpha(k)$ grows and
hits a Landau pole at a finite value of $k^2$ that we call
$\Lambda^{2}_{L}$, where $\alpha(\Lambda_{L})\ra\infty\,.$ With an
initial condition fixed in region I, there is no solution to the
renormalization group equation when $k^2\geq \Lambda^{2}_{L}\,.$ Yet
the integral over the euclidean momentum in eq.~(\ref{eq:momentum})
generated by perturbation theory runs up to $\infty$. It has been
recently shown by Grunberg~\cite{Gr95} that in the presence of the
first two terms in the power expansion of the
$\beta$--function it is still possible to find an exact change of
variables which maps $k^2$ integrals in the euclidean space, like the
one  in (\ref{eq:momentum}), to integrals in the
$z$--variable of the Borel plane.  The change of variables in question
is 

\be
\label{eq:grunberg}
B\frac{z}{2}=\frac{1-\frac{\alpha(m)}{\alpha(k)}}{1+\delta
B\alpha(m)}\,,
\ee
\noi
and leads to the Borel sum

\be \label{eq:borel2}
a_{e}=\frac{1}{2\pi}\int_{0}^{\infty}dz\,e^{-\frac{z}{\alpha}}
\frac{e^{-z\delta B}} {(1-\frac{Bz}{2})^{1+\delta}}\,.
\ee
\noi
However this is only a formal expression. With the initial condition
$\alpha(m)>0$ fixed, the change of variables in (\ref{eq:grunberg}) is
only well defined in the region where
$m^2\leq k^2 <\Lambda^{2}_{L}\,,$ which corresponds to
$0\leq B\frac{z}{2}< (1+\delta B\alpha(m))^{-1}<1\,.$ Let us however
inspect what happens when one goes beyond this  physically
allowed region i.e., when $(1+\delta B\alpha(m))^{-1}\leq
B\frac{z}{2}<\infty$. This is equivalent to taking a certain arbitrary 
$\alpha(k)$ in the
non--physical region III, where
$(1+\delta B\alpha(m))^{-1}<B\frac{z}{2}\leq 1$, and then in the region
II, where $1\leq B\frac{z}{2} <\infty\,.$ With this proviso, eq.
(\ref{eq:borel2}) has a definite (but arbitrary) meaning. From this point of
view, the singularity at $z=2/B$ acts as a reminder of this arbitrariness. 
For example, one can define an
``extended'' $\alpha(k)$ valid at all $k^2$--values, $m^2\leq
k^{2}\leq\infty$, such that in the physical region $m^2\leq
k^{2}<\Lambda_{L}^{2}$  it coincides with the physical
$\alpha(k)$ and that it satisfies the perturbative renormalization group
equation in (\ref{eq:rgegs}), except at a finite number of
singularities. This extended $\alpha(k)$ has the following 
form\footnote{Notice the absolute value in the
argument of the logarithm, in contrast to eq. (\ref{eq:rgegs}).}:


\be\label{eq:rgers}
\frac{\alpha(m)}{\alpha(k)}=1-B\alpha(m)\log\frac{k^2}{m^2}+\delta
B\alpha(m)
\log\left|\frac{\alpha(m)\left(1+\delta B\alpha(k)\right)}
{\alpha(k)\left(1+\delta B\alpha(m)\right)}
\right|\,.
\ee
\noi
This expression leads to the trajectory with the arrows displayed
on Fig.~2, where the arrows correspond to the flow as one integrates
over $z$ going first from region I to region III, and then to region
II. In so doing, one sees that one crosses three singularities~: one
in which $\alpha(k)$ goes to infinity --the conventionally called
Landau singularity-- and two in which $\log\frac{k^2}{m^2}$ goes to
infinity --at the fixed points of $\beta(\alpha)$~: $\alpha^{*}=0$ and
$\alpha^{**}=-\frac{1}{\delta B}\,.$ With this ``extended'' definition of
$\alpha(k)$, the change of variables in (\ref{eq:grunberg}) leads to
the Borel sum

\be\label{eq:nbi}
a_{e}=\frac{1}{2\pi}\int_{0}^{\infty}dz\,e^{-\frac{z}{\alpha}}
\frac{e^{-z\delta B}} {(1-\frac{Bz}{2})\left|(1-\frac{Bz}{2})
\right|^{\delta}}\,.
\ee
\noi
The two loop Borel integrals (\ref{eq:borel2}) and (\ref{eq:nbi}) still have 
the singularity at
$z=2/B$, as in the one-loop case. This may lead one to believe that, exactly
as in the one-loop case, this singularity is caused by the corresponding
Landau singularity in euclidean momentum. However this is not so. 
To see this explicitly let us
expand eq.~(\ref{eq:rgegs}), or eq.~(\ref{eq:rgers}), for 
$\alpha(k)\ra +\infty$ i.e. near the Landau pole~:

\be
B\alpha\log\frac{k^2}{m^2}=1-\delta
B\alpha\log\left(1+\frac{1}{\delta
B\alpha}\right)-\frac{\alpha}{2\delta B}\frac{1}{\alpha^{2}(k)}
+\cO\left(\frac{1}{\alpha^{3}(k)}\right)\,.
\ee 
\noi
The terms of $\cO\left(\frac{1}{\alpha(k)}\right)$ cancel, which
means that the Landau pole of $\alpha(k)$ at

\be
k^{2}=\Lambda^{2}_{L}=m^{2}\exp{\left\{\frac{1}
{B\alpha}-\delta\log\left(1+\frac{1}{\delta
B\alpha}\right) \right\}}\,,
\ee
\noi
goes like

\be
\alpha(k)=\frac{\pi}{\beta_{2}^{1/2}}\,\frac{1}
{(\log\Lambda^{2}_{L}-\log k^{2})^{1/2}}\,;
\ee  
\noi
i.e., it is a squared root singularity and therefore {\it integrable} up to
the Landau scale $\Lambda_L^2$ where $Bz/2=(1+\delta B \alpha(m))^{-1}<1$. 
 From that point onwards the integration in $z$ can be done using 
the extended
$\alpha(k)$ in eq. (12), following the arrows on Fig. 2 from region I to
region III until $\alpha=\alpha^{**}$ where one hits the Borel singularity 
at $z=z_n$. This
is to be contrasted with the one loop behaviour in eq.~(\ref{eq:ech1})
where

\be
\alpha(k)=\frac{1}{B}\,\frac{1}{\log\Lambda^{2}_{L}-\log k^{2}}\,,
\ee
\noi
and the singularity is not integrable. 
The euclidean origin of the $z=2/B$ singularity in
the Borel sum at the two--loop level is the $\log\frac{k^2}{m^2}$
singularity corresponding to $\alpha^{**}=-\frac{1}{\delta B}\,;$ not the
Landau pole~! Of course in the limit $\beta_2\to 0$ \footnote{Although one
can invoke $1/N_F$ arguments to effect the limit $\beta_2\to 0$ in QED,
there is no analogue in the case of QCD.} one recovers the one-loop
situation in which $\alpha^{**}\to \infty$, and the Landau pole.
\vspace*{7mm}

{\bf 4.} The basic features we have discussed in the case of the
anomalous magnetic moment of the electron in QED are rather generic and
they appear as well in the case of IR--renormalon calculus in QCD
where, typically, one encounters euclidean integrals like

\be
R_{n}(Q^2)=\int_{0}^{Q^2}\frac{dk^2}{k^2}\left(\frac{k^2}{Q^2}\right)^{n}\,
\alpha_{s}\left(\frac{k}{Q},\alpha_{s}(Q)\right)\,,
\ee
\noi with $n>0$, so that the integral is infrared convergent.
The scale $Q^2$ in these integrals corresponds to a sufficiently
large choice of euclidean momentum at which the QCD running coupling
$\alpha_{s}(Q)$ is reasonably small. Here the
ambiguity problem appears because of the integration over virtual
euclidean momentum in the infrared region $0\leq k^2 \leq Q^2$, where
the extrapolation of the perturbative $\alpha_{s}(Q)$ coupling is not
well defined. It is often stated in the literature that the reason for
the {\em ambiguity} is the existence of the Landau pole in the region
of integration. We shall see that, as already shown in the case of
$a_{e}$ in QED, this is only correct if one rather arbitrarily
restricts the QCD beta function to its first term.

The discussion runs parallel to the one in the previous section, except
that now the running coupling
$\alpha_{s}\left(\frac{k}{Q},\alpha_{s}(Q)\right)\equiv
\alpha_{s}(k)$ obeys the QCD renormalization group equation

\be \label{eq:rgeqcd}
\frac{d\alpha_{s}(k)}{d\log\frac{k^2}{Q^2}}
=-b_{0}\alpha_{s}^{2}(k)-b_{1}\alpha_{s}^{3}(k)\,,
\ee
\noi where we are using the notation

\bea b_{0} \equiv   -\frac{1}{2\pi}\,\beta_{1}^{\mbox{\footnotesize{\rm
{}~QCD}}} & = &
\frac{1}{12\pi}(11N_{c}-2n_{f})\,, \\ b_{1}  \equiv 
-\frac{1}{2\pi^{2}}\,\beta_{2}^{\mbox{\footnotesize{\rm ~QCD}}} &  = &
\frac{1}{8\pi^2}\left\{\frac{17}{3}N_{c}^{2}-
\frac{N_c^2-1}{2N_c}n_{f}-
\frac{5}{3}N_cn_f \right\}\,.
\eea
with $b_0,b_1>0$ in the physically interesting case $N_c=n_f=3$. 
We can consider eq.~(\ref{eq:rgeqcd}) to be a generic case since any
other
$\beta$--function can be brought into this form by an appropriate
coupling constant redefinition. Of course this redefinition
will entail in general an infinite power series. The solutions of
eq.~(\ref{eq:rgeqcd}) have the pattern shown in Fig.~4, with region I
corresponding to the physical region with the boundary condition
$\alpha_{s}(Q)\geq 0$. The equivalent change of variables to the one  in
eq.~(\ref{eq:grunberg}) is \cite{Gr95}

\be \label{eq:grunbergn}
\frac{z}{z_{n}}=\frac{1-\frac{\alpha_{s}(Q)}{\alpha_{s}(k)}}
{1+\frac{b_{1}}{b_{0}}\alpha_{s}(Q)}\,; \qquad z_{n}\equiv 
\frac{n}{b_{0}}\,.
\ee
\noi It maps the $k^2$ integral $R_{n}$ onto an integral in
the $z$--variable of the Borel plane~:

\be \label{eq:birn} R_{n}(Q^2)=-\frac{1}{n}\int_{0}^{\infty} dz
e^{-\frac{z}{\alpha_{s}(Q)}}
\frac{e^{-z\frac{b_{1}}{b_{0}}}}
{\left(1-\frac{z}{z_{n}}\right)^{1+\delta_{n}}}\,, \qquad 
\delta_{n}\equiv n\frac{b_{1}}{b_{0}^{2}}\,.
\ee

For a fixed $n$, the Borel integral (\ref{eq:birn}) 
is singular at $z=z_{n}$, which according to (\ref{eq:grunbergn})
happens when $\alpha_{s}(k)=-\frac{b_{0}}{b_{1}}$ and has nothing to do
with the Landau pole at which, by definition,
$\alpha_{s}(k)=\infty\,.$ In fact at the Landau pole, which now occurs
when

\be k^2=\Lambda_{L}^2=Q^2\exp{\left\{\frac{-1}{b_{0}\alpha_{s}(Q)}
+\frac{b_{1}}{b_{0}^2}\log\left( 1 +
\frac{b_0}{b_1 \alpha_{s}(Q)}\right)\right\}}\,,
\ee
\noi
$\alpha_{s}(k)\ra\infty$ as a squared root singularity~:

\be
\alpha_{s}(k)=\frac{1}{(2b_{1})^{1/2}}\,\frac{1}{(\log\Lambda^{2}_{L}-\log
k^{2})^{1/2}}\,,
\ee
\noi and it is therefore integrable.

Notice that the ambiguity in eq. (\ref{eq:birn}) can be parameterized by a
power--like contribution in $Q^2$. Indeed, making the change of variables 
\be \frac{z}{z_n}=1+\frac{\alpha(Q) \omega}{z_n (1+\frac{b_1}{b_0}
\alpha(Q))}
\ee
one finds that the ambiguity in $R_n(Q^2)$, $\delta R_n(Q^2)$, is given by
\be \delta R_n(Q)= \left\{\frac{z_n^{1+\delta_n}}{n}
e^{-z_n\frac{b_1}{b_0}} \int_* 
\frac{d\omega \ e^{-\omega}}{\omega^{1+\delta_n}}\right\} 
\left(\frac{1}{\alpha(Q)} +
\frac{b_1}{b_0}\right)^{\delta_n} e^{-\frac{z_n}{\alpha(Q)}}
\ee     

\be \label{eq:ambiguity} 
\delta R_n(Q^2)= \{{\rm coefficient}\} 
\times \left(\frac{\Lambda_L^2}{Q^2}\right)^n
\ee
where $\int_* d\omega \ e^{-\omega}\ \omega^{-(1+\delta_n)}$ denotes an
integral with an arbitrary prescription to skip the singularity at
$\omega=0$. Although, as repeatedly emphasized, the Landau singularity is
not directly responsible for the singularity at $z=z_n$, the final result
of eq. (\ref{eq:ambiguity}) is insensitive to this fact and the form of 
(\ref{eq:ambiguity}) is the same as in the case of the one-loop $\beta$
function.

Before we conclude, we would like just to mention that an analogous
reasoning for $a_e$ in QED leads, mutatis mutandis, to an equation like
(\ref{eq:ambiguity}) for the analogous quantity $\delta a_e$ but with $n=-2$
and $Q^2=m^2$.  Therefore it is also true for a two-loop $\beta$ function
that the ambiguity $\delta a_e$ can be given by the tree-level insertion of
the local operator  $\left(m^3/\Lambda^4_L\right)
\bar{\psi}(x)\sigma_{\mu\nu} \psi(x) F^{\mu\nu}(x)$, just as in the one-loop
case.
 
\vspace{7 mm} {\bf Acknowledgments~:}
 
S.P. thanks E. Bagan and A. Calsina for clarifying conversations. 
He also thanks T. Coarasa and E. Bagan for their help with the graphics.  
E.de R. has benefited from the ``de Betancourt
-- Perronet'' prize for visits to Barcelona. We thank R. Tarrach for a
critical reading of the manuscript.
 
\vskip 2cm

Note Added:

The analysis of renormalon effects in the case of a $\beta$ function 
with a perturbative fixed point has been considered by Grunberg in 
ref. [21] and also by Yu.L. Dokshitzer and N.G. Uraltsev in 
hep-ph/9512407. They reach similar conslusions to ours. We thank N.G.
Uraltsev for drawing his paper to our attention.

\newpage

 
\vspace{4cm}
 
\noi {\Large \bf Figure Captions}
 
\vspace{7mm}

\noi Fig. 1\,: Feynman diagrams which define the one--renormalon chain
contribution to $a_{e}$ in eq.~(\ref{eq:ane}).
\vspace{3mm}

\noi Fig. 2\,: General pattern of the QED renormalization group solution
in eq.~(\ref{eq:rgegs}). $\tau$ stands for $\log \frac{k^2}{m^2}$. 
The regions I, II and III correspond to
the choice $\alpha(m)\geq 0$;
$-\frac{1}{\delta B}\leq\alpha(m)\leq 0$; and $\alpha(m)\leq
-\frac{1}{\delta B}$ as boundary condition.  
\vspace{3mm}

\noi Fig. 3\,: The QED $\beta$--function in perturbation theory at the
two--loop level. The arrows show the flow of the effective charge as
$k^2\ra\infty$.
\vspace{3mm}

\noi Fig. 4\,: General pattern of the QCD renormalization group solution
in eq.~(\ref{eq:rgeqcd}). $\tau$ stands for  $\log \frac{k^2}{Q^2}$.
The regions I, II and III correspond to the
choice $\alpha_{s}(Q)\geq 0$;
$-\frac{b_{0}}{b_{1}}\leq\alpha_{s}(Q)\leq 0$; and $\alpha_{s}(Q)\leq
-\frac{b_{0}}{b_{1}}$ as boundary condition.  
\vspace{3mm}


\begin{thebibliography}{99}
 
\bibitem{Dy52}
        F.J. Dyson, Phys. Rev. {\bf 85}Ê(1952) 631.

\bibitem{LeGZJ90}
        J.C. Le Guillou and J. Zinn-Justin, {\it Large--Order
        Behaviour of Perturbation Theory}, Current Physices Sources and
        Comments, Vol.~7, North--Holland (1980)

\bibitem{tH78}
        G. 't Hooft, in {\it The Whys of Subnuclear Physics}, Erice
        1977, ed. A. Zichichi, Plenum, New York 1977.

\bibitem{La77}
        B. Lautrup, Phys. Lett. {\bf 69B} (1977) 109.

\bibitem{Pa77}
        G. Parisi, Phys. Lett. {\bf 76B}Ê(1977) 65.

\bibitem{Pa79}
        G. Parisi, Nucl. Phys. {\bf B150} (1979) 163.

\bibitem{Co81}
        R. Coquereaux, Phys. Rev. {\bf D23}Ê(1981) 2276.

\bibitem{BS95}
        M. Beneke and V.A. Smirnov, {\it Ultraviolet Renormalons in
Abelian Gauge Theories.} Preprint hep-ph/9510437.

\bibitem{Da82}
        F. David, Nucl. Phys. {\bf B209} (1982) 465.

\bibitem{Da84}
        F. David, Nucl. Phys. {\bf B234}Ê(1984) 493.


\bibitem{NSVZ84}
        V.A. Novikov, M.A. Shifman, A.I. Vainshtein, and V.I.
        Zakharov, Phys. Reports {\bf 116} (1984) 105.


\bibitem{NSVZ85}
        V.A. Novikov, M.A. Shifman, A.I. Vainshtein, and V.I.
        Zakharov, Nucl. Phys. {\bf B249} (1985) 445.


\bibitem{Mu85}
        A.H. Mueller, Nucl. Phys. {\bf B250} (1985) 327.


\bibitem{Da86}
        F. David, Nucl. Phys.{\bf B263}Ê(1986) 508.



\bibitem{Mu93}
        A.H. Mueller, in {\it QCD--Twenty Years Later}, World
        Scientific, Singapore, 1993.

\bibitem{Sj95}
        C.T. Sachrajda, {\it Renormalons}. Preprint hep-lat/9509085. 

\bibitem{BY92}
        L.S. Brown, L.G. Yaffe, and Ch. Zhai, Phys. Rev. {\bf D46}
        (1992) 4712.

\bibitem{Za92}
        V.I. Zakharov, Nucl. Phys. {\bf B385} (1992) 452.

\bibitem{VZ94a}
        A.I. Vainshtein and V.I. Zakharov, Phy. Rev. Lett. {\bf 73}
        (1994) 1207.

\bibitem{VZ94b}
        A.I. Vainshtein and V.I. Zakharov, {\it
        Ultraviolet--Renormalon Reexamined.} Preprint hep-ph/9404248.
      


\bibitem{Gr95}
        G. Grunberg, {\it Renormalons and Perturbative Fixed
        Points.} Preprint hep-ph/9511435. 




     
   
       

\end{thebibliography}
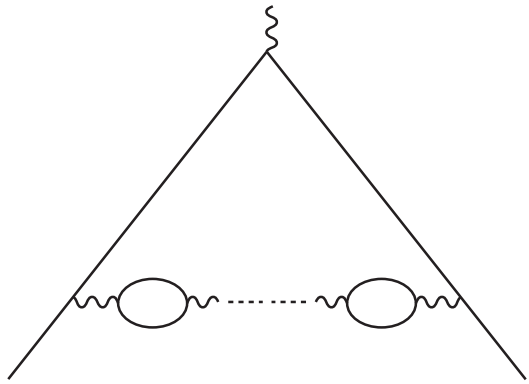
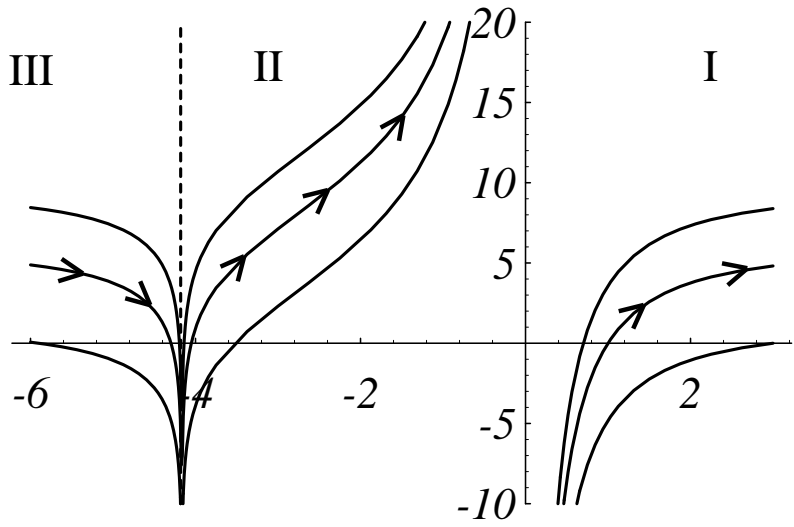
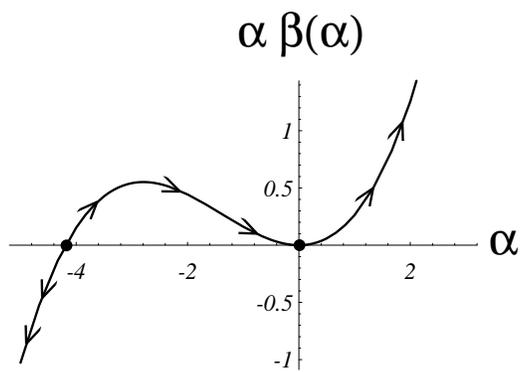
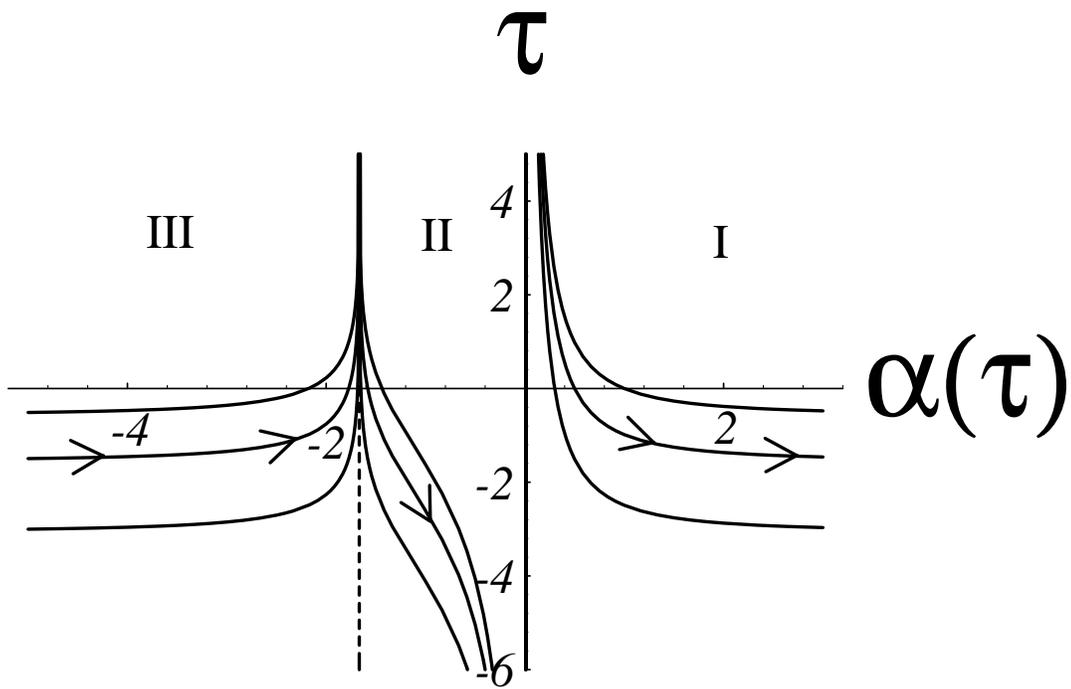
\end{document}